\documentstyle[epsf,psfig,sprocl]{article}

\def\Journal#1#2#3#4{{#1} {\bf #2}, #3 (#4)}

\def\NIMA{{\em Nucl. Instrum. Methods} A}
\def\NPB{{\em Nucl. Phys.} B}

\def\PRD{{\em Phys. Rev.} D}
\def\ZPC{{\em Z. Phys.} C}

\def\ra{\rightarrow}

\def\be{\begin{equation}}
\def\ee{\end{equation}}
\def\bea{\begin{eqnarray}}
\def\eea{\end{eqnarray}}
\def\nub{{\overline{\nu}}}

\begin{document}

\title{MEASUREMENT OF $\sin^2 \theta_W$ IN $\nu N$ SCATTERING FROM NuTeV
}

\author{ R.~A.~JOHNSON$^{1}$, T.~ADAMS$^{4}$, {A.~ALTON$^{4}$}, 
{S.~AVVAKUMOV$^{7}$},
 L.~DE~BARBARO$^{5}$, P.~DE~BARBARO$^{7}$, R.~H.~BERNSTEIN$^{3}$,
 A.~BODEK$^{7}$, T.~BOLTON$^{4}$, J.~BRAU$^{6}$, D.~BUCHHOLZ$^{5}$,
 H.~BUDD$^{7}$, L.~BUGEL$^{3}$, J.~CONRAD$^{2}$, R.~B.~DRUCKER$^{6}$,
 R.~FREY$^{6}$, {J.~GOLDMAN$^{4}$}, {M.~GONCHAROV$^{4}$},
 D.~A.~HARRIS$^{7}$, S.~KOUTSOLIOTAS$^{2,*}$,
 {J.~H.~KIM$^{2,\alpha}$}, M.~J.~LAMM$^{3}$, W.~MARSH$^{3}$, 
 {D.~MASON$^{6}$}, {C.~MCNULTY$^{2}$}, K.~S.~MCFARLAND$^{7}$, 
 D.~NAPLES$^{4,+}$, P.~NIENABER$^{3}$, {A.~ROMOSAN$^{2}$}, 
 W.~K.~SAKUMOTO$^{7}$,
 H.~SCHELLMAN$^{5}$, M.~H.~SHAEVITZ$^{2}$, P.~SPENTZOURIS$^{2}$$^{**}$ , 
 E.~G.~STERN$^{2,\dagger}$, {M.~VAKILI$^{1,\parallel}$}, {A.~VAITAITIS$^{2}$}, 
 {V.~WU$^{1}$}, {U.~K.~YANG$^{7}$}, J.~YU$^{3}$, and 
 {G.~P.~ZELLER$^{5}$}
}

\address{
 $^{1}$University of Cincinnati, Cincinnati, OH 45221,             
 $^{2}$Columbia University, New York, NY 10027,                   
 $^{3}$Fermi National Accelerator Laboratory, Batavia, IL 60510,  
 $^{4}$Kansas State University, Manhattan, KS 66506,              
 $^{5}$Northwestern University, Evanston, IL 60208,               
 $^{6}$University of Oregon, Eugene, OR 97403,                 
 $^{7}$University of Rochester, Rochester, NY 14627 \\         
\hspace{2ex} \\             
 Current Addresses:
 $^{*}$Bucknell University,
 $^{\ddagger}$Lawrence Berkeley National Laboratory, 
 $^{\parallel}$Texas A\&M University,
 $^{**}$Fermi National Accelerator Laboratory,
 $^{+}$University of Pittsburgh,
 $^\alpha$University of California at Irvine,
 $^\dagger$Lucent Technologies
} 

\address{Presented by Randy Johnson}

\maketitle\abstracts{ We report the measurement of $\sin^2\theta_w$ in $\nu N$ 
deep inelastic scattering from the NuTeV experiment.  By using separate 
neutrino and anti-neutrino beams, NuTeV is able to determine $\sin^2\theta_W$
with smaller systematic and similar or smaller statistical errors 
when compared to
previous neutrino experiments.
NuTeV measures 
${\sin^2\theta_W}^{\rm (on-shell)}=0.2253
\pm 0.0019{\rm (stat.)} \pm 0.0010 {\rm (sys.)}$, 
which implies 
$M_W = 80.26 \pm 0.11$ GeV/c$^2$.
}

\section{Introduction}

Neutrino experiments play a key role in establishing the validity of the
electroweak Standard Model.  Even today with the large samples of $W$ and $Z$
boson events, precision $\nu N$ experiments are still competitive in 
determining the electroweak model parameters and also show that the model
is valid over many orders of magnitude in $q^2$.  To date, neutrino experiments
have used the Llewellyn Smith\cite{smith} relationship to determine 
$\sin^2 \theta_W$.\cite{CCFR,others}  
However, the dominate systematic error, the experimental determination
of the effective charm mass, has limited the precision of these measurements.  
Paschos and Wolfenstein\cite{paschos} have shown that if the neutral and
charged current cross sections for neutrinos and anti-neutrinos could be 
measured separately, then the ratio
\bea
R^- &=& \frac{\sigma(\nu_\mu N \ra \nu_\mu X)-\sigma(\nub_\mu N \ra \nub X)}
{\sigma(\nu_\mu N \ra \mu^- X)-\sigma(\nub_\mu N \ra \mu^+ X)} 
\nonumber \\
&=& \frac{R^\nu - r R^\nub}{1-r} = \frac{1}{2} - \sin^2\theta_W
=g_L^2-g_R^2
\label{eq:rm}
\eea
(where $R^\nu = 
\frac{\sigma(\nu_\mu N \ra \nu_\mu X)}{\sigma(\nu_\mu N \ra \mu^- X)}$,
$R^\nub = 
\frac{\sigma(\nub_\mu N \ra \nu_\mu X)}{\sigma(\nub_\mu N \ra \mu^+ X)}$, 
$r = 
\frac{\sigma(\nub_\mu N \ra \mu^+ X)}{\sigma(\nu_\mu N \ra \mu^- X)}$,
$g_{L,R}^2 = u^2_{L,R}+d^2_{L,R}=$ the sum of the squares of the 
quark couplings, and the relationship to $\sin^2\theta_W$ is a tree 
level relationship only)
eliminates the effect of the sea quarks on the determination of $\sin^2 
\theta_W$ and, hence, the majority of the error from the mass of the 
charm quark.  To make such a measurement was a goal of the NuTeV experiment
at Fermilab.

\section{NuTeV experiment}

\begin{figure}[htb]
\centerline{
\epsfxsize=2.25in
\epsfbox{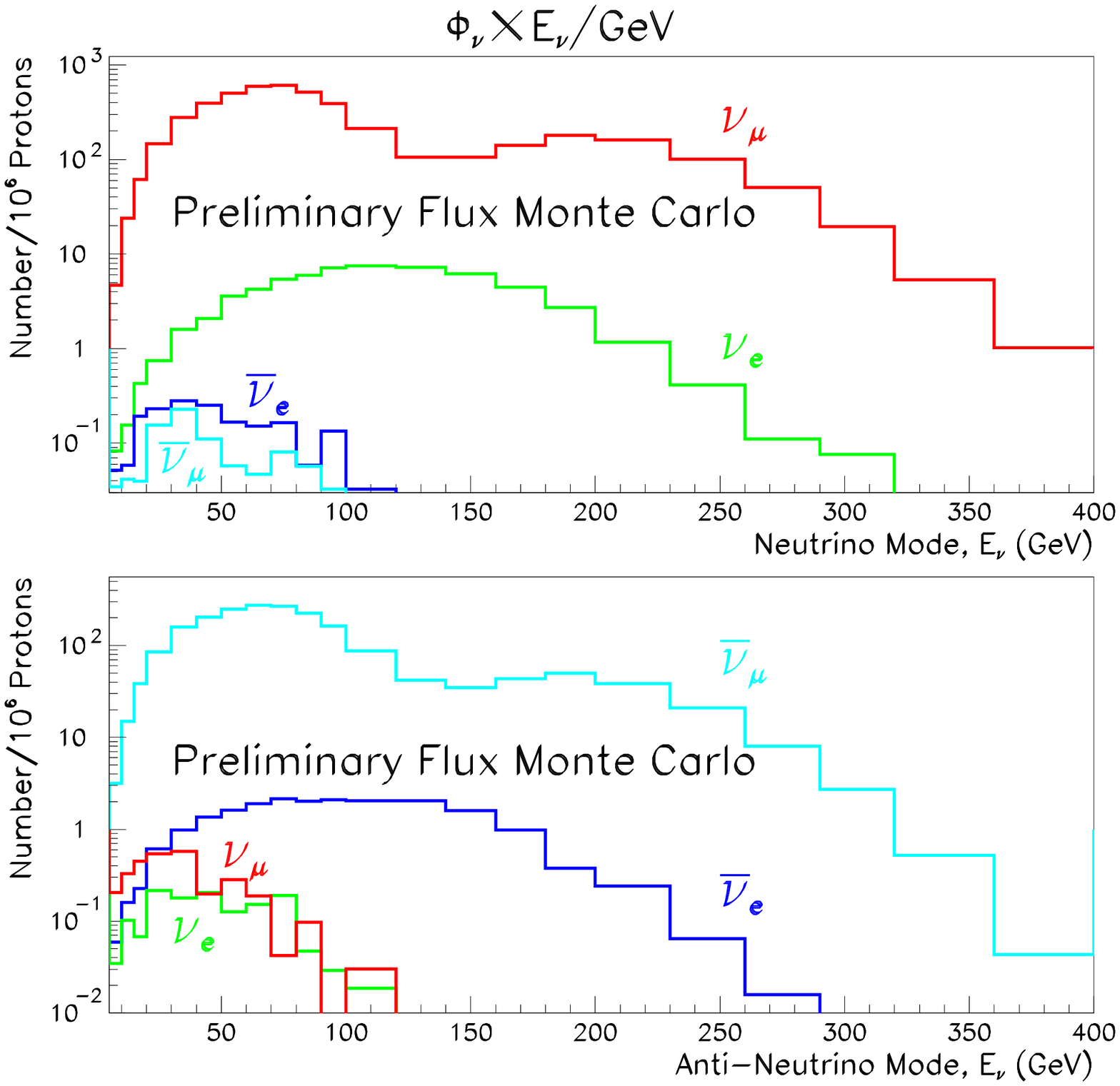}
\hfill
\epsfxsize=2.25in
\epsfbox{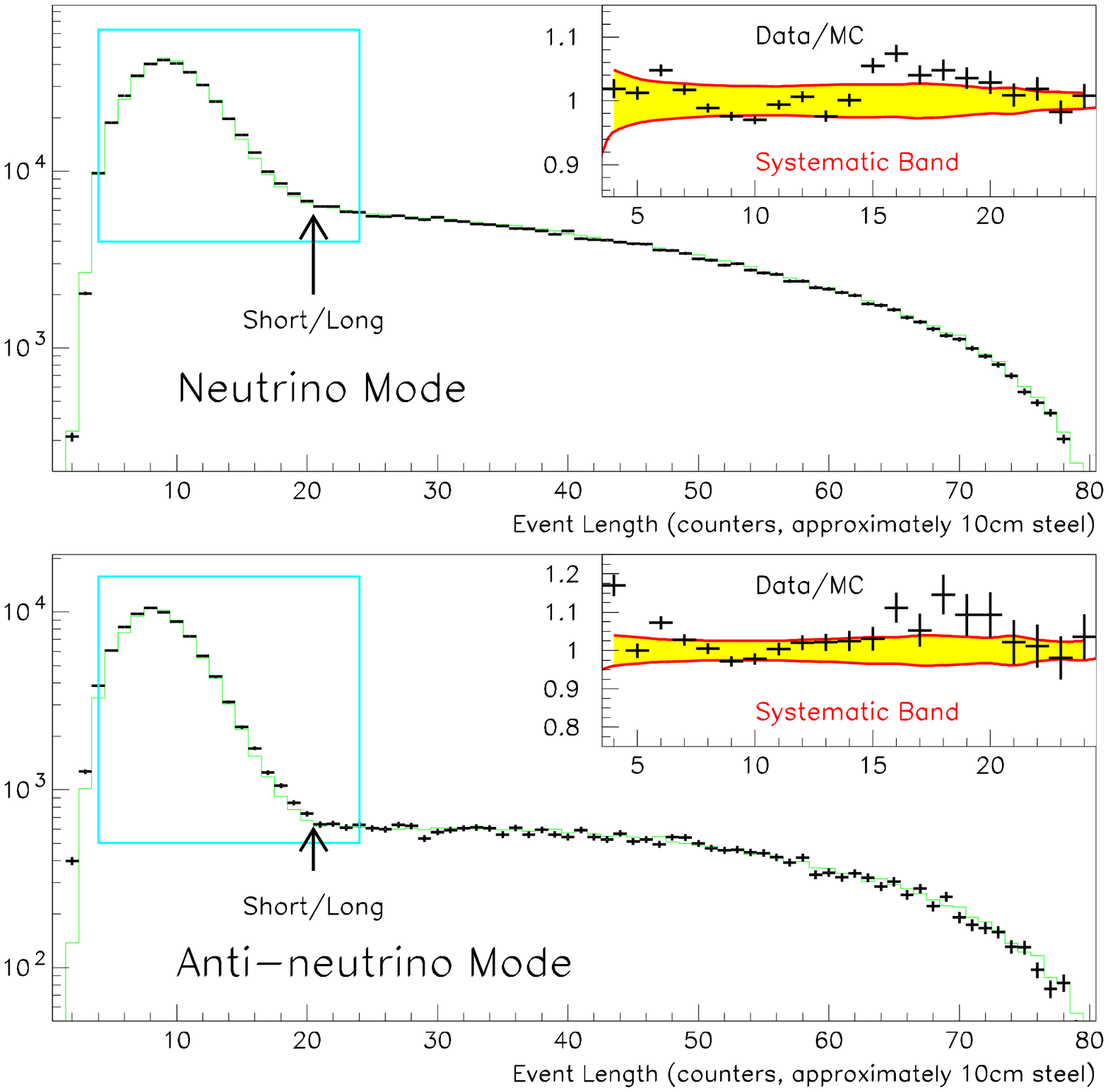}
}
\centerline{\hspace*{1in}(a)\hspace*{1in}
\hfill
\hspace*{1in}(b)\hspace*{1in}}
\caption{(a) {\em Preliminary} NuTeV Neutrino and Anti-Neutrino Fluxes;
(b) {\em Preliminary}  NuTeV Event Length Distribution. 
\label{fig:lengthandflux}}
\end{figure}

In order to measure $R^-$, the cross sections for $\nu N$ and $\nub N$
scattering must be measured separately.  For NuTeV, we designed and built a 
Sign Selected Quadrapole Train (SSQT) which selected the charge of
the mesons that were directed into our decay volume, and, thus, selected
whether neutrinos or anti-neutrinos hit the detector.  The fluxes of neutrino
varieties in neutrino and anti-neutrino modes are shown in 
Fig.~\ref{fig:lengthandflux}(a).
The wrong sign contamination fraction was less than $1\times 10^{-3}$ in
neutrino mode and $2\times10^{-3}$ in anti-neutrino mode.  The electron
neutrino contamination in the beam is a serious background in the
determination of $R^-$.  The majority of $\nu_e$'s come from charged
kaon decays which can be kinematically related to the high energy $\nu_\mu$
spectrum that comes from the same particles.  The design of the SSQT
greatly reduced the contribution of the neutral kaon decays to the 
$\nu_e$ flux in comparison to other experiments. 


The NuTeV experiment used the refurbished CCFR iron calorimeter and toroid
spectrometer that is described elsewhere.\cite{NIM}  
Data were collected during the '96-'97 Fermilab 
fixed target run.  Charged current
$\nu$ and $\nub$ events were differentiated from neutral current events
by the length in the calorimeter over which energy was deposited.  
Long events indicate the presence of a muon and are classified as
charged current events; neutral current events only contain a hadronic
shower and are therefore short. 
The length
distribution for neutrino and anti-neutrino events is shown in 
Fig.~\ref{fig:lengthandflux}(b).   A detailed Monte Carlo program was used to
determine the contamination of neutral current events in the long distribution
and of charged current events in the short distribution.  The detector 
parameters used in this program were measured using a test beam which 
constantly monitored and calibrated the detector.


\section{Results}

Rather than measuring $R^-$ directly, we chose to measure a ``pseudo'' $R^-$
\be
{\rm pseudo}R^- = R^\nu - \alpha R^\nub \label{eq:pseudoR}
\ee
where $\alpha = 0.5136$ was chosen to minimize the effects of the charmed
quark mass on our result given the differences in fluxes and cross 
sections between the $\nu$ and $\nub$ modes.  We varied the value of 
$\sin^2\theta_W$ in our Monte Carlo until the predicted value of pseudo$R^-$ matched
the measured one.

The preliminary result from the NuTeV data is
\bea
&&\sin^2{\theta_W}^{\rm (on-shell)} =  0.2253
\pm 0.0019{\rm (stat.)} \pm 0.0010 {\rm (sys.)} \nonumber \\
&&-0.00142\times\left(\frac{M_{\rm top}^2 - (175~{\rm GeV/c}^2)^2}
{(100 {\rm~GeV/c}^2)^2}
\right)
+0.00048 \times \log_e \left(\frac{{M_{\rm Higgs}}^2}{150 {\rm~GeV/c}^2}
\right)\!.~~~
\label{eq:result}
\eea
The small residual dependence of our result on $M_{\rm top}$ and $M_{\rm 
Higgs}$ comes from the leading terms in the electroweak radiative 
corrections\cite{radiative}.  The sources of the statistical and 
systematic errors for this result are given in Table~\ref{tab:errors}.

\begin{table}[t]
\caption{
Sources of uncertainties in the determination of $\sin^2\theta_W$.
\label{tab:errors}}
\vspace{0.2cm}
\begin{center}
\footnotesize
\begin{tabular}{|r|c|}
\hline
  SOURCE OF UNCERTAINTY & $\delta\sin^2\theta_W$ \\
\hline \hline
  { {\em Statistics}: \hfill {Data} }  & {0.00188}  \\
  { Monte Carlo } & {0.00028} \\ \hline \hline
  { \bf TOTAL STATISTICS \hfill } & {0.00190} \\ \hline\hline
  { {\em $\nu_e/\nub_e$ Flux}: \hfill $K^\pm$ ($1.1\%$) } & {0.00024} \\
  { Other sources of $\nu_e$'s } & {0.00048} \\
  { {\em Energy Measurement}: \hfill  Calibrations ($0.5\%$)} & {0.00043} \\
  { Muon Energy Deposition ($3\%$)} & {0.00004} \\
  { Energy Resolution } & {0.00004} \\
  { {\em Event Length}:  \hfill Hadron Shower } & {0.00015} \\
  { Longitudinal Vertex Determination } & {0.00015} \\
  { Counter Edge Location } & {0.00010} \\
  { Counter Efficiency and Noise } & {0.00016} \\
  \hline
  { \bf TOTAL EXP. SYST. \hfill } & {0.00075} \\ \hline\hline
  { {\em Sea Quarks}: \hfill Strange Sea } & {0.00034}  \\
  { $V_{cd}$ } & {0.00004} \\
  { Charm Sea } & {0.00009} \\
  { Charm Mass} & {0.00009} \\
  { {\em Other $\nu/\nub$ Cross-Section Differences}: \hfill 
       $\sigma^\nu/\sigma^{\nub}$ } & {0.00023}  \\
  { Non-Isoscalar Target } & {0.00013} \\
  { Radiative Corrections } & {0.00051} \\ 
  { {\em Non-QPM Cross-Section}: \hfill Higher Twist } & {0.00013} \\ 
  { Longitudinal Structure Function } & {0.00007}  \\ \hline
  { \bf TOTAL PHYSICS MODEL \hfill } & {0.00070} \\  \hline \hline 
  { \bf TOTAL UNCERTAINTY \hfill } & {0.0022} \\  \hline
\end{tabular}
\end{center}
\end{table}

\begin{figure}[htb]
\centerline{
\hfill
\epsfxsize=1.7in
\epsfbox[50 40 460 540]{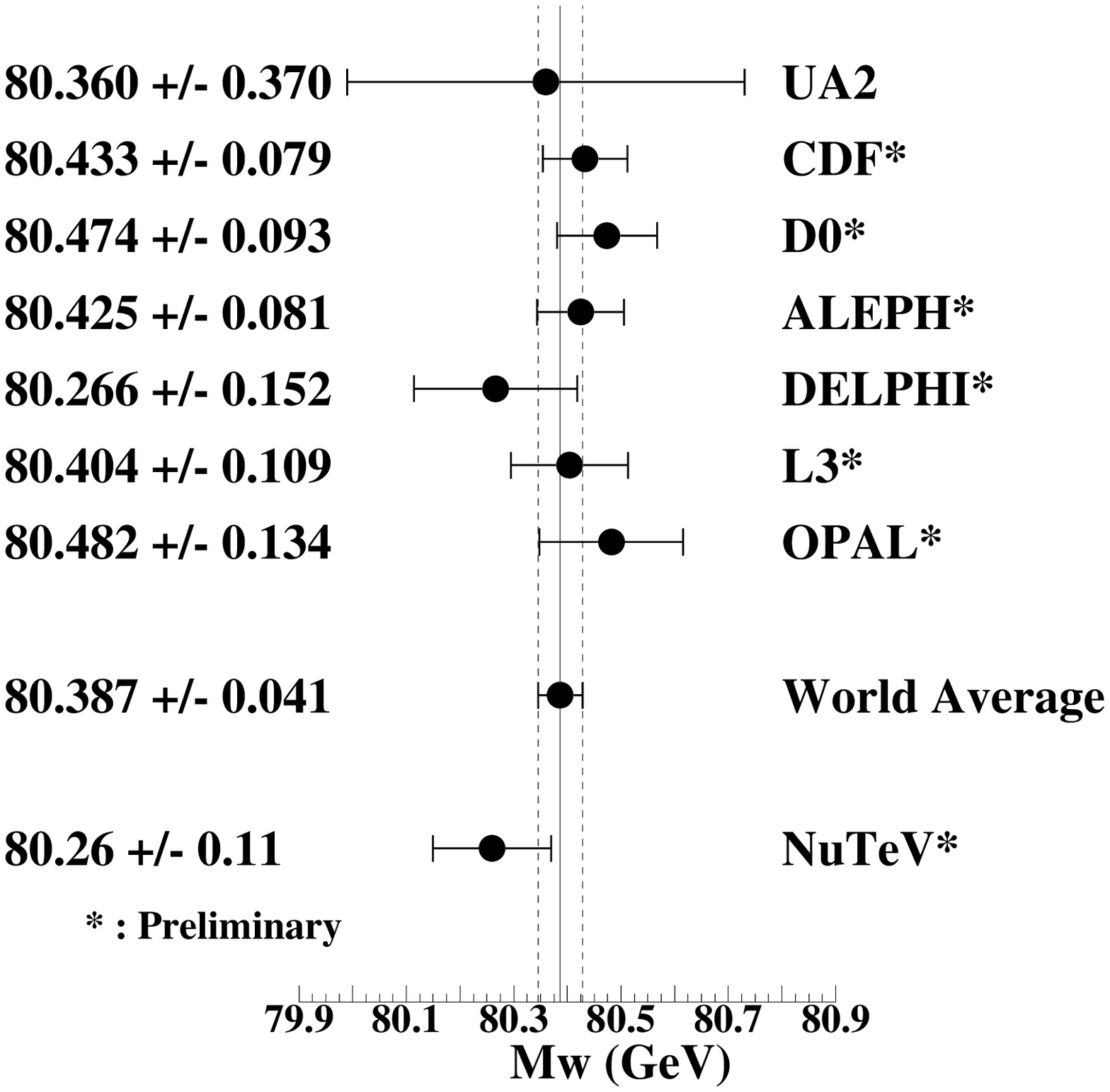}
\hfill
\epsfxsize=1.7in
\epsfbox[50 40 460 540]{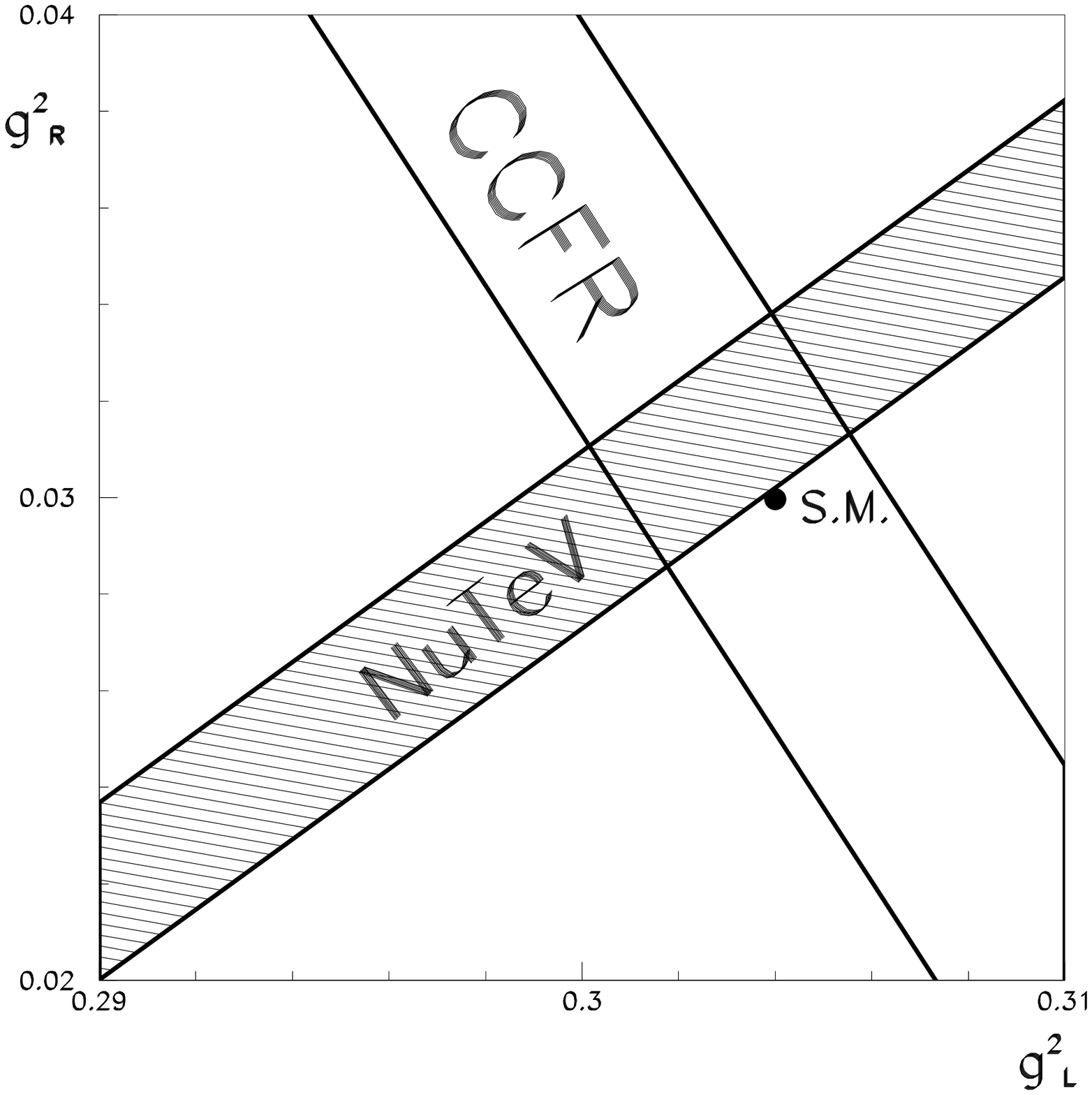}
\hfill
}
\centerline{\hfill\hspace*{.75in}(a)\hspace*{.75in}\hfill\hspace*{.75in}
(b)\hspace*{.75in}\hfill}
\caption{(a) Comparison of the measured $M_W$ mass from NuTeV and other experiments; (b) Comparison of the model independent couplings between 
NuTeV and CCFR (Point is the '98 Standard Model fit of world data by the 
Particle Data Group$^7$).
\label{fig:newmwandcouplings}}
\end{figure}

Since $\sin^2{\theta_W}^{\rm (on-shell)} = 1-M_W^2/M_Z^2$, the result 
given in Eq.~\ref{eq:result} is equivalent to 
\bea
M_W &&= 80.26 \pm 0.10({\rm stat.}) \pm 0.05({\rm sys.}) \nonumber \\
&&+0.073\times\left(\frac{M_{\rm top}^2 - (175~{\rm GeV/c}^2)^2}
{(100 {\rm~GeV/c}^2)^2}
\right)
-0.025 \times \log_e \left(\frac{{M_{\rm Higgs}}^2}{150 {\rm~GeV/c}^2}
\right)\!.~~~
\label{eq:mw}
\eea
This result is compared with other measurements of the $W$ mass in 
Fig.~\ref{fig:newmwandcouplings}(a).


It is possible to extract the NuTeV result in a model independent framework
in terms of the left and right handed quark couplings.  For our data
\bea
0.4530 - \sin^2\theta_W &=& 0.2277\pm 0.0022 \nonumber \\
&=& 0.8587 u_L^2 + 0.8828 d_L^2 -1.1657 u_R^2 -1.2288 d_R^2.
\label{eq:coupling}
\eea
(Note the similarity of this with Eq.~\ref{eq:rm}.)  
This result is plotted on the $g^2_L$-$g^2_R$ plane in 
Fig.~\ref{fig:newmwandcouplings}(b) and compared with the CCFR
result which measured $R^\nu = g^2_L + 0.64 g^2_R$.
The results of the two experiments 
are consistent with the Standard Model.


\section*{Acknowledgments}
We gratefully acknowledge the substantial contributions in the construction
and operation of the NuTeV beamlines and the refurbishment of the 
NuTeV detector made by the staff of the Fermilab Beams and Particle 
Physics Division.  This work was supported by grants from the U.S. Department
of Energy, from the National Science Foundation, and from the Alfred P. Sloan 
Foundation.

\end{document}